\documentclass[a4paper, 10pt]{article}  % Comment this line out
\usepackage{graphicx} % for pdf, bitmapped graphics files
\usepackage{amsmath} % assumes amsmath package installed
\usepackage{amssymb}  % assumes amsmath package installed
\usepackage{subcaption}
\usepackage{cite}
\usepackage{amsfonts}
\usepackage{graphicx}
\usepackage{xcolor}
\usepackage{csvsimple}
\usepackage{tikz} 
\tikzset{every picture/.style={line width=0.75pt}} %set default line width to 0.75pt    
\usepackage{pgfplots}
\usepackage{booktabs}
\usepgfplotslibrary{dateplot}

\newtheorem{remark}{Remark}

\title{\LARGE \bf
Combating District Heating Bottlenecks Using Load Control
}

\author{Felix Agner$^{\star1}$, Pauline Kergus$^{\star}$, Richard Pates$^{\star}$ and Anders Rantzer$^{\star}$% <-this % stops a space
}

\date{}

\newcommand*{\review}{\textcolor{black}}

\newcommand*{\rad}{\text{hs}}
\newcommand*{\ext}{\text{ext}}
\newcommand*{\iin}{\text{in}}
\newcommand*{\out}{\text{out}}

\newcommand*{\hs}{\text{hs}}
\newcommand*{\degC}{^\circ\mathrm{C}}

\newcommand*{\q}{\mathbf{q}}

\begin{document}

\maketitle
\thispagestyle{empty}
\pagestyle{empty}

\footnotetext{$^{\star}$Department of Automatic Control, Lund University, Sweden}
\footnotetext{$^{1}$Contact: \tt\small felix.agner@control.lth.se}
\footnotetext{This work is funded by the European Research Council (ERC) under the European Union's Horizon 2020 research and innovation program under grant agreement No 834142 (ScalableControl).}

%%%%%%%%%%%%%%%%%%%%%%%%%%%%%%%%%%%%%%%%%%%%%%%%%%%%%%%%%%%%%%%%%%%%%%%%%%%%%%%%
\begin{abstract}
\iffalse % Old abstract part
In current district heating systems, central distribution pumps are used to drive a flow through the network. If the heat load is too great, hydraulical bottlenecks cause peripheral units (customers) to experience reduced flow rates. This work aims to increase fairness through a coordinating control strategy. 
\fi

\review{The 4th generation of district heating systems face a potential problem where lowered water temperatures lead to higher flow rates, which requires higher hydraulic capacity in terms of pipe and pump sizes. This increases the effect of the already existing issue of hydraulic bottlenecks, causing peripheral units (customers) to experience reduced flow rates. A coordinating control strategy is presented in this work aimed at reducing the effect of such bottlenecks on the comfort of customers.}
This is done by distributing the flow deficit over many units rather than a few. Previous works mainly focus on MPC-structured controllers that depend on complex system models and online optimization techniques. This work proposes a method that requires little information about models for individual units and minimal IT communication between control systems. The proposed method is compared with a traditional control strategy and an optimal baseline in a simulation study. This shows that the proposed method can decrease the worst case indoor temperature deviations.

\end{abstract}

%%%%%%%%%%%%%%%%%%%%%%%%%%%%%%%%%%%%%%%%%%%%%%%%%%%%%%%%%%%%%%%%%%%%%%%%%%%%%%%%

\section{INTRODUCTION}

%% LAYOUT OF INTRODUCTION

% Describe general problems in DH

% Describe the problem of lacking pressure

% Describe previous knowledge and adaptations to DSM

% Describe knowledge of modelling hydraulic networks

% Contributions of this work

% Structure of article

\begin{figure*}[]
    \centering
    \begin{tikzpicture}
        \newcommand{\pw}{5.5cm}
        \newcommand{\ph}{4.5cm}
        \node (a) at (0,0)
         {
            \begin{tikzpicture}
				\begin{axis}[
					width=\pw, height=\ph,
					xmin = 0, xmax = 200,
					ymin = 16, ymax = 21,
					grid = both,
					xlabel = {Time [h]},
					ylabel = {\textcolor{red}{Indoor Temperature [$\degC$]}},
					y tick label style = {color=red}
					]
					\addplot[red] table[col sep = comma,x=time,y=1]{Data/Tin_g.csv};
				\end{axis}
				\begin{axis}[
                    width=\pw, height=\ph,
                    xmin = 0, xmax = 200,
                    ymin = -20, ymax = 5,
                    hide x axis,
                    axis y line*=right,
                    ylabel near ticks
					]
					\addplot[dotted] table[col sep = comma,x=time,y=Tout]{Data/Tout.csv};
				\end{axis}
			\end{tikzpicture}
         };
         \node (c) at (a.east) [xshift = 5cm]
         {
            \begin{tikzpicture}
				\begin{axis}[
					width=\pw, height=\ph,
					xmin = 0, xmax = 200,
					ymin = 16, ymax = 21,
					grid = both,
					xlabel = {Time [h]},
					ylabel = {\textcolor{red}{Indoor Temperature [$\degC$]}},
					y tick label style = {color=red}
					]
					\addplot[red] table[col sep = comma,x=time,y=23]{Data/Tin_g.csv};
				\end{axis}
				\begin{axis}[
                    width=\pw, height=\ph,
                    xmin = 0, xmax = 200,
                    ymin = -20, ymax = 5,
                    hide x axis,
                    axis y line*=right,
                    ylabel = {Outdoor Temperature [$\degC$]},
                    ylabel near ticks
					]
					\addplot[dotted] table[col sep = comma,x=time,y=Tout]{Data/Tout.csv};
				\end{axis}
			\end{tikzpicture}
         };
        \node (b) at (a.south) [anchor=north,yshift=0cm,xshift=4.0cm]
         {
            \scalebox{.9}{\input{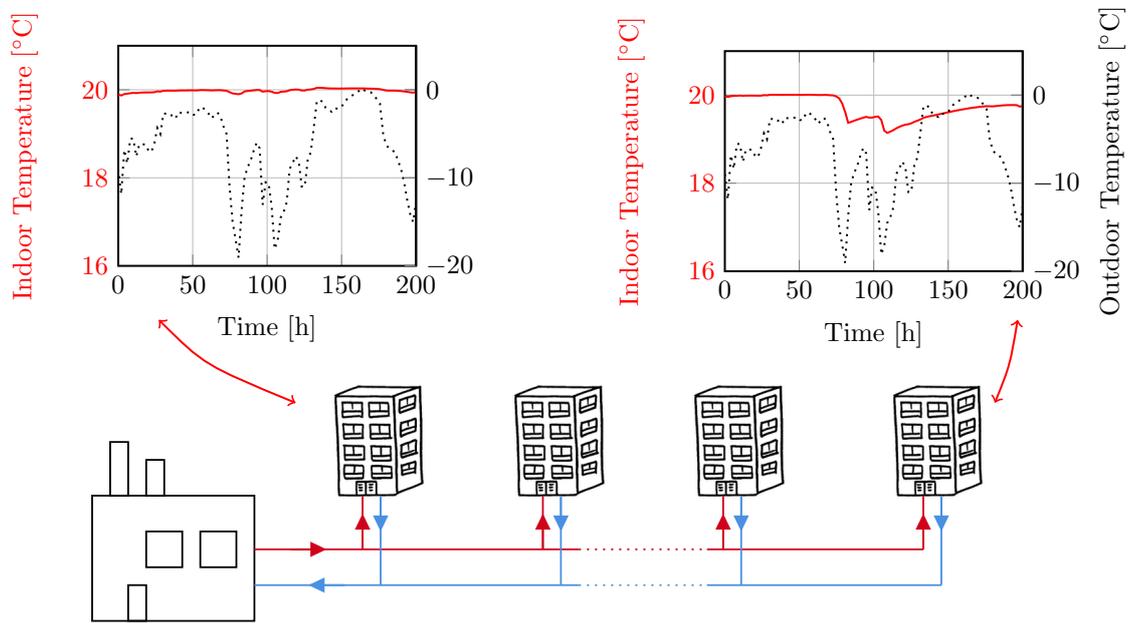}}
         };
         \draw [color=red, <->, ]   (-1.1,-1.9) .. controls (-0.5,-2.5).. (0.7,-3) ;
         \draw [color=red, <->, ]   (10.2,-1.9) .. controls (10.1,-2.5).. (9.9,-3) ;
    \end{tikzpicture}
    \caption{Discrepancy in indoor temperature (red) between units connected to the grid. When the outdoor temperature (dotted) becomes critically cold, units close to the pressure source are able to maintain indoor comfort temperature while units far from the source are not.}
    \label{fig:introduction figure}
\end{figure*}

\review{One important puzzle piece of the smart energy system of the future is the integration of a variety of energy sources and distribution methods. This allows harnessing synergies and reducing the impact of stochastic fluctuations in energy supply and demand \cite{smartenergyfor100renewable}.  District heating systems have been shown to be a powerful tool in this energy system, but research indicates that a transformation of district heating from the old 3rd generation to a new 4th generation is needed. An important characteristic of this emerging 4th that allows it to be integrated into the overall energy chain is reduced supply water temperatures which would allow using previously untapped heat sources such as renewable sources and industrial waste heat \cite{lund_4th_2014}. In theory, the reduction in supply temperature should be accompanied by an equal drop in return temperature, leading to an equal temperature difference and thus no alternation in the necessary flow \cite{statusof4g}. However, lowering the building return temperatures requires an improvement in space heating technology \cite{comparisonoflowtemp}, \cite{statusof4g} and if such is not the case there may be a reduction in differential temperature. This leads to higher flows needed to distribute the same amount of power, implying that the piping and pumping power of 4th generation district heating systems may have to be dimensioned for higher capacity. This presents an additional cost. It also reduces the potential of retrofitting existing infrastructure for lower grid temperatures, which otherwise may prove a cost-effective solution \cite{statusof4g}, \cite{brange_decision-making_2019}. If the grid capacity is not dimensioned for higher flows, it may lead to \textbf{bottlenecks} \cite{brange_decision-making_2019}.}

% Old introduction
\iffalse Indoor climate control plays a considerable part in global energy consumption \cite{perez2008review} and in areas such as in northern Europe, district heating is an important part of this puzzle. An important characteristic of the emerging 4th generation of district heating systems is reduced supply water temperatures which would allow using previously untapped heat sources such as renewable sources and waste heat \cite{lund_4th_2014}. In some parts of the world, district heating systems are already well developed, and it would therefore be favorable to convert already existing infrastructure to lower supply temperatures \cite{brange_decision-making_2019}. However to compensate  for the reduced supply temperatures, higher flow rates are needed which can lead to the phenomenon of \textbf{bottlenecks} \cite{brange_decision-making_2019}.\fi %End of old introduction
Bottlenecks imply that some part of the network can experience severe drops in differential pressure. Buildings (units) connected in these parts may find it hard to extract sufficient flows to keep indoor temperatures at comfort level. \review{In fact, this is not only a hypothetical problem in future generations of district heating but is} already a problem in currently operating networks \cite{thebible}. This phenomenon arises under \textbf{peak load conditions}, i.e. when the flows in the network are high, coinciding with when the outdoor temperature is low. Figure \ref{fig:introduction figure} shows this problem, based on simulation which will be explained later in this article. When the outdoor temperature becomes too low, the indoor temperatures start to differ from each other. Buildings close to the pressure source maintain comfort temperature while it becomes cold in buildings further down the distribution line. Reducing the effect of bottlenecks could increase \textbf{robustness} to low outdoor temperatures, in the sense that a drop in outdoor temperature would cause reduced worst-case deviations in indoor temperature. Apart from the possibility of reducing supply temperatures, this could also grant the possibility of extending existing networks, and designing larger new networks with less concern for critical outdoor temperature and the influence this would have on customer comfort. One approach to tackling this issue is through the use of demand side management, as suggested in \cite{thebible} \cite{vandermeulen_controlling_2018} \cite{brange_decision-making_2019} \cite{guelpa_demand_2021}.
\\

Demand side management is an umbrella term for different ways of altering the demand of customers connected to a grid. There is a rich history of demand side management in the power grid literature, but it has also begun making an appearance in the district heating literature \cite{guelpa_demand_2021}. This article focuses specifically on direct load control, i.e. directly altering and deciding the heat load of customers. A common approach to direct load control is to use \textbf{centralized} optimization with the objective of minimizing some operational cost for the entire network. For instance an optimization scheme was introduced in \cite{bhattacharya_demand_2019} to improve fairness of heat distribution in a line network. As the optimization problems tend to grow drastically with the number of connected units, these methods can run into problems of scalability. Another approach is to have \textbf{decentralized} optimal controllers, such as in \cite{saletti_development_2020}, and then combine their signals to compute desired heat production. However in this scenario there is no coordination between the different units to ensure that they request a load that is feasible. 
\\

To ensure that any enacted heat loads are within the system constraints, a good model of the grid is needed. In practice, the considered distribution model depends on the purpose of the model as well as the design of the network, as summarized in \cite{sarbu_review_2019}. Some works choose to disregard aspects of the constraints given by the network such as pressure losses or time delays \cite{bhattacharya_demand_2019}. One common approach to model flows and pressures of the network is to assume that the specific heat load at each unit has to be met, and then using these loads to calculate the flows realized in the network \cite{hydraulic_optim_meshed}, \cite{benny_operational_nodate}, \cite{larsen_aggregated_2002}. However this approach does not hold in the case where the desired heat loads of each building cannot be met due to the corresponding flows being too large for the network to handle. To the authors' knowledge, there is little work on describing the limits on the flows in the system.
\\

This work is focused on understanding how bottlenecks can be combated through direct load control, in such a way that the hydraulic constraints of the network are taken into account and the control structure remains scalable for a large number of connected units. \review{The idea is to combine the increased ease of implementation of distributed controllers with the system-level benefits of a centralized strategy. An important distinction to make is that many works on demand side management try to optimize some operational cost, e.g. cost of energy production units. In this work we assume that the system operates at full capacity, and the objective is simply to distribute the supplied energy fairly between customers.}

The contribution of this work is in three parts;
\begin{enumerate}
    \item formulating the constraints limiting the unit flows in a line-structured district heating grid;
    \item introducing a load coordination scheme that builds on the traditional control architecture of district heating such that it should be easy to implement in existing networks, and;
    \item comparing two control architectures; traditional control and the aforementioned load coordination architecture, with an optimal baseline reference through a simulation study.
\end{enumerate}

The work is presented as follows: Section \ref{sec:problem formulation} introduces a mathematical formulation of the problem. The notion of robustness to outdoor temperatures is introduced here. Section \ref{sec:system models} presents the mathematical models of the network and the units connected to the network. Section \ref{sec:control architectures} defines two different control architectures and an optimal baseline reference, which are then compared in a simulation study described in section \ref{sec:simulation and results}. The results and future work are finally discussed in section \ref{sec:summary}.

%%%%%%%%%%%%%%%%%%%%%%%%%%%%%%%%%%%%%%%%%%%%%%%%%%%%%%%%%%%%%%%%%%%%%%%%%%%%%%%%

% \subsection*{List of notation}

\begin{table*}[]
    \centering
    \caption{Table of notation used in this article.}
    \label{tab:notation}
    \begin{tabular}{|c|l|l|} \toprule
        Symbol & Description & Unit \\ \midrule
        $J$ & Cost function related to discomfort & $\degC$s\\
        $T$ & Temperature & $\degC$\\
        $T_c$ & Comfort temperature & $\degC$ \\
        $T_\iin$ & Indoor temperature & $\degC$ \\
        $e$ & Difference between indoor and comfort temperature & $\degC$ \\
        $T_\hs$ & Temperature of water in heating system & $\degC$ \\
        $T_\ext$ & Outdoor temperature & $\degC$ \\
        $T_\text{sup}$ & Primary side supply temperature & $\degC$ \\
        $T_\text{ret}$ & Primary side return temperature & $\degC$ \\
        $t$ & Time & s \\
        $t_s$ & Sampling time & s \\
        $q$ & Water flow & kg/s \\
        $\mathcal{Q}$ & The set of admissible system water flows & - \\
        $C_\iin$ & Heat capacity of indoor area & J/$\degC$ \\
        $C_\hs$ & Heat capacity of water in radiator systems & J/$\degC$ \\
        $C_w$ & Specific heat capacity of water & J/kg$\degC$ \\
        $R_\ext$ & Heat resistance between building interior and exterior & W/$\degC$\\
        $R_\hs$ & Heat resistance between radiator system and building interior & W/$\degC$\\
        $P$ & Furnished heat power & \\
        $A,B_q,B_\ext$ & Matrices defining the dynamics of simulated buildings & - \\
        $\Delta p$& Differential pressure & Pa \\
        $\mathcal{L}$ & Network loop & - \\
        $F$ & Network incidence matrix & - \\
        $a$ & Hydraulic resistance & Pa/(kg/s)$^2$\\
        $c$ & Pump curve parameters & - \\
        $r$ & Pump frequency ratio & - \\
        $\alpha_0$,$\alpha_1$ & Heating system temperature set-point parameters & - \\
        $k$ & Building P-controller gain & kg/s$\degC$ \\
        $\delta$ & Flow set-point deviation & kg/s \\
        $\gamma$ & Coordination weight factor & s$\degC$/kg\\
        $\lambda$ & Coordination price factor & - \\
        \bottomrule 
    \end{tabular}
\end{table*}

%%%%%%%%%%%%%%%%%%%%%%%%%%%%%%%%%%%%%%%%%%%%%%%%%%%%%%%%%%%%%%%%%%%%%%%%%%%%%%%%

\section{PROBLEM AND SYSTEM FORMULATION}\label{sec:system models}

This section formalizes the problem of this work. Part \ref{sec:problem formulation} puts the problem to be solved in mathematical form. Part \ref{sec:buildings} presents the model of building temperature dynamics and the union between buildings and the district heating grid. Part \ref{sec:network} explains the hydraulic model of the distribution network, dictating the constraints on hot water flow in the system. Formulating these constraints in closed form constitutes the first contribution of the article. 

\subsection{Problem Formulation}\label{sec:problem formulation}

The control problem of this work is to maintain comfortable indoor temperatures in all buildings connected to a district heating network even under extreme disturbances in the form of low outdoor temperatures. The control signal deciding the amount of heat furnished to each building $i$ is the flow of hot water $q_i$ through their substation. Mathematically, the problem is formulated as:

\begin{subequations}
\begin{alignat}{3}
&\!\min_{\mathbf{q}(t_k)}        &\qquad& J(\mathbf{T}) \label{eq:problem formulation cost}\\
&\text{subject to} &      & \mathbf{T}(t_{k+1}) = f(\mathbf{T}(t_{k}),\mathbf{q}(t_k), T_\ext(t_k)) \label{eq:problem formulation dynamics}\\
&                  &      & \mathbf{q}(t_k) \in \mathcal{Q}.\label{eq:problem formulation flow constraint}
\end{alignat}
\end{subequations}
What this means is that we want to minimize some discomfort $J$ related to the indoor temperatures $\mathbf{T}$ in the connected buildings. These temperatures $\mathbf{T}$ evolve according to dynamics $f$, which depend on the furnished flows $\mathbf{q}$ and the outdoor temperature $T_\ext$. This relationship $f$ will be detailed in the next section, \ref{sec:buildings} and is in this work modelled linearly. Lastly, the furnished flow $\mathbf{q}$ is limited by the capacity of the distribution system. The set $\mathcal{Q}$ of flows that can be realized in the system is the subject of section \ref{sec:network}.
\\

The cost function $J$ should capture the discomfort experienced by each customer. To define this discomfort, consider the temperature deviation $e_i(t_k)$ for each unit $i$ connected to the grid at each point in time $t_k$. $e_i(t_k)$ is the deviation between the desired comfort temperature $T_{c,i}$ and the actual indoor temperature $T_{\iin,i}(t_k)$.
\begin{equation}
    e_i(t_k) = T_{c,i}-T_{\iin,i}(t_k)
    \label{eq:temperature error}
\end{equation}
The discomfort $J_i$ experienced by a unit during a time period $t = t_1,t_2,\hdots, t_K$ can then be defined as
\begin{equation}
    J_i = \sum_{k=1}^K |e_i(t_k)t_s|,\label{eq:discomfort}
\end{equation}
where $t_s$ is the time in between times $t_k$ and $t_{k+1}$. Note also three candidates for measuring the system-level discomfort, $J_1$, $J_2$ and $J_\infty$:
\begin{align}
    J_1 &= \sum_{k=0}^K  \frac{1}{N}\sum_{i=1}^N |e_i(t_k)t_s |\label{eq:cost 1norm}\\ 
    J_2 &= \sum_{k=0}^K \sqrt{\frac{1}{N^2}\sum_{i=1}^N |e_i(t_k)t_s |^2} \label{eq:cost 2norm}\\
    J_\infty &= \sum_{k=0}^K \underset{i}{\text{max}} (|e_i(t_k)t_s |) \label{eq:cost infnorm}
\end{align}
Here $J_1$ is a metric for the sum of discomfort experienced by all units, $J_2$ is a metric for the total discomfort where larger units discomfort are penalized more, and $J_\infty$ is a metric for the worst discomfort experienced in the grid. The scenario we want to avoid is for extreme discomfort levels to arise in any unit, and for this reason the $J_\infty$-cost is the cost that will be used in the controller design of section \ref{sec:control architectures}. The two remaining costs, $J_1$ and $J_2$ will be used for evaluation as a point of reference.
\\

\remark

Some works also consider optimizing over the power required to actuate the flows and temperatures in the grid and thus minimize the cost of running the system. For instance \cite{saletti_development_2020} consider the utilized pumping power and \cite{bacher_identifying_2011} consider the electrical heating power in an adjacent problem considering an electrically heated unit. In this work we don't consider the cost of running the system. As we are interested in fair distribution under extremely cold situations, it is assumed that the heat and pumping power supplied to the system will have to be at maximum capacity. The interest is rather in understanding how to distribute this supplied power between connected units.

\subsection{Buildings}\label{sec:buildings}

Here we investigate the dynamics dictating the temperatures in each building, i.e. the function $f$ of (\ref{eq:problem formulation dynamics}). With each building $i$, we associate two states $T_{\iin ,i}$ and $T_{\hs ,i}$, representing the mean indoor temperature and mean temperature of heating system circulating water respectively. This allows the construction of the following state space representation:

\begin{align}
    C_{\iin ,i}\dot{T}_{\iin ,i} &= -(\frac{1}{R_{\ext ,i}}+\frac{1}{R_{\rad ,i}})T_{\iin ,i} + \frac{1}{R_{\rad ,i}}T_{\rad, i} + \frac{1}{R_{\ext ,i}}T_\ext \label{eq:Tin} \\
    C_{\rad ,i}\dot{T}_{\rad ,i} &= \frac{1}{R_{\rad ,i}}T_{\iin ,i} - \frac{1}{R_{\rad ,i}} T_{\hs ,i} + P_{i} \label{eq:Trad},
\end{align}
where $C_{\iin ,i}$ and $C_{\rad ,i}$ is heat capacity of the building interior and heating system respectively. $P_{i}$ is the heat power extracted from the primary side of the district heating system. The heat energy flow between interior and exterior as well as between heating and system interior are proportional to the inverse of the heat resistances $R_{\ext ,i}$ and $R_{\rad ,i}$ respectively. These types of models of varying complexity have been used extensively in literature on modeling building temperature dynamics,\cite{bacher_identifying_2011}, \cite{bhattacharya_demand_2019}, \cite{saletti_development_2020}, and can be augmented to capture different levels of complexity. In this work, a simple model of buildings is used, motivated by the interest in understanding the general distribution of temperatures in a large set of buildings, rather than the details of one individual building. The presented continuous time state space representation can then be transformed into a discrete time representation of the system if a standard zero-order-hold assumption is made for the inputs $T_\ext$ and $P_i$.
\\

The heat energy, $P_i$, extracted from the network is here assumed to be proportional to the water flow through the primary side pipes of the building substation and the temperature difference between supply and return pipes in the network, $(T_\text{sup} - T_\text{ret})$:

\begin{equation}
    P_i = C_w (T_\text{sup} - T_\text{ret}) q_i
    \label{eq:P}
\end{equation}

where $C_w$ is the \review{specific} heat capacity of water.\review{ In the simulations and analysis in this work, the supply and return temperatures in the network are considered constant. This simplification is made to simplify simulations and analysis. While these temperatures are not constant in a real system, they are measured in building substations. As such, they could be included in the control strategy, where the now constant values would simply be exchanged for measured values.}

% \iffalse
% %OLD
% Here it is assumed that $T_\text{sup} - T_\text{ret}$ is constant over the entire network which is a simplification used to facilitate analysis. In reality the supply temperature $T_\text{sup}$ varies with time and space in the network. However the simplification can be motivated by that in certain conditions the heat generated through pressure losses in the pipes may counteract the heat losses, leading to an almost constant supply temperature \cite{thebible}. The return temperature $T_\text{ret}$ at each substation depends on the flows on both sides of the substation heat exchanger, the temperature of the water in the secondary side of the heat exchanger and the design of the heat exchanger itself. \cite{gustafsson_improved_2010}
% \fi
% \\

To simplify the equations above, we can gather the indoor temperatures of all buildings into one vector $\mathbf{T}$, and the dynamics can then be put on the following linear form:

\begin{equation}
    \mathbf{T}(t_{k+1}) = A\mathbf{T}(t_{k}) + B_q \mathbf{q}(t_k) + B_\ext T_\ext(t_k)
    \label{eq:temp_dynamics}
\end{equation}

\remark
\review{In this model we do not take domestic hot water use into account, much due to the difficulty of including a realistic model of this usage. In a typical scenario, the flow $q_i$ through each building would consist of two parts, one for space heating and one for hot water usage. This is one aspect that should be considered in future work. It could either be included as another part of the control system, or modeled as a disturbance on the system.}

\subsection{Distribution Model}\label{sec:network}

\begin{figure}[h!]
	\centering
	\scalebox{.6}{\input{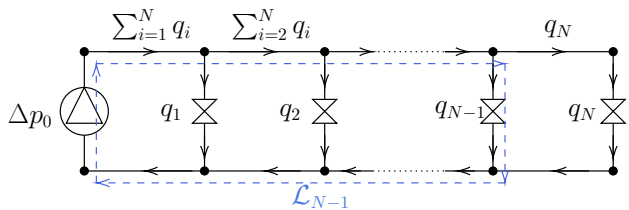} }
	\caption{Simple network structure with only one heat source. Here equation (\ref{eq:flow equality}) has already been used to calculate the flows in the supply and return pipes as a function of the substation flows $q_i$. Loop $N-1$ is illustrated with blue arrows.}	
	\label{fig:convex network}
\end{figure}

This part formulates the constraints on hot water flows in the distribution network, the first contribution of this article. This corresponds to the set $\mathcal{Q}$ of equation (\ref{eq:problem formulation flow constraint}). This work considers primarily a simple network architecture corresponding to a line of $N$ units, as seen in Figure \ref{fig:convex network}. A central pump circulates the water through the pipes, and each substation, with index $i$, has a control valve that it can use to regulate the water flow $q_i$ through their substation locally.
Associate with each pipe and valve a hydraulic resistance $a_i(t)$. $a_i(t)$ is constant for pipes and variable for valves, but bounded below by $a_i(t) \geq a_i^\text{min}$ corresponding to a completely open valve. Note that the hydraulic resistance of supply pipes will be denoted $a_{i}^{\text{sup}}$ and for return pipes $a_i^{\text{ret}}$.

The following equations dictate the relation between flows and pressure head in the network. For a pipe or valve $i$,

\begin{equation}
	\Delta p_i = a_i(t) q_i^2, 
	\label{eq:pipe}
\end{equation}
where $\Delta p_i$ is the pressure difference between the entrance and exit points of the component, caused by pressure losses due to friction \cite{thebible}, \cite{hydraulic_optim_meshed}. For a pump $j$, 
\begin{equation}
    \Delta p_j = c_{1,j} q_j^2 + c_{2,j} r_j(t) + c_{3,j} r_j(t)^2
    \label{eq:pump curve}
\end{equation}
where $c_{1,j}$, $c_{2,j}$ and $c_{3,j}$ are pump parameters that denote the characteristics of a specific pump, and $r_j(t) \leq 1$ is the pump frequency ratio indicating the capacity at which the pump is operating at. 
\\

Two laws apply to the flows and pressures in the network \cite{hydraulic_optim_meshed}:

\begin{enumerate}
    \item The sum of directed flows entering a node is 0, so that the volume of water in a specific node does not change. \label{item:constraint1}
    \item Traversing a loop of pipes in the network results in a 0 net change in pressure. \label{item:constraint2}
\end{enumerate}
\ref{item:constraint1}) can be expressed as
\begin{equation}
	F\mathbf{q}(t) = 0,
	\label{eq:flow equality}
\end{equation}
where $F$ is the incidence matrix of the network. \review{The incidence matrix defines how all the pipes in the grid are connected to nodes (connection points) in the network and is defined as}

\begin{equation}
    F_{ij} = \begin{cases}
    1, \quad \text{pipe j leads to node i} \\
    -1,\quad \text{pipe j leads away from node i} \\
    0, \quad \text{pipe j is not connected to node i}
    \end{cases}
\label{eq:incidence}
\end{equation}

When applied to the network in Figure \ref{fig:convex network}, we see that the flows in the supply-and-return pipes can be expressed as sums of the substation flows $q_i$. The second constraint \ref{item:constraint2}) can be expressed as 
\begin{equation}
	\sum_{i\in\mathcal{L}_l} \Delta p_i = 0,
	\label{eq:pressure equality}
\end{equation}
where $\mathcal{L}_l$ denotes the $l$th loop in the network, and $\Delta p_i$ is the pressure difference along each edge that constitutes that loop \cite{SULZERPUMPS201027}, \cite{hydraulic_optim_meshed}. In Figure \ref{fig:convex network} we can identify $N$ loops. Loop $l$ starts in the central pump, goes through the supply pipes with resistances $a_i^\text{sup}$, then through the valve of substation $l$ with resistance $a_l(t)$, and then back through the return pipes with resistances $a_i^\text{ret}$. The net pressure difference along this loop is then
\begin{equation}
\begin{split}
    q_l(t)^2a_l(t) + \sum_{j=1}^l (a_j^\text{sup} + a_j^\text{ret})(\sum_{i=j}^N q_i(t))^2 \\
    = c_1(\sum_{i=1}^N q_i(t))^2 + c_2r_(t) + c_3r(t)^2,
\end{split}
    \label{eq:loop net pressure}
\end{equation}
where the left expression is the pressure losses in the pipes and the right expression is the pressure head generated by the pump. There are $N$ constraints on this form, one for each loop $l$ for $l= 1\dots N$. Any flow $q$ that satisfies the inequality:
\begin{equation}
    \begin{split}
	 q_l(t)^2a_l^{\text{min}} + \sum_{j=1}^l (a_j^\text{sup} + a_j^\text{ret})(\sum_{i=j}^N q_i(t))^2 \\ \leq c_1(\sum_{i=1}^N q_i(t))^2 + c_2 + c_3        
    \end{split}
	\label{eq:convex inequality}
\end{equation}
 can also be made to satisfy the equation (\ref{eq:loop net pressure}), by choosing a significantly large $r(t)$ and a sufficiently large $a_l(t)$. Therefore any flow that satisfies (\ref{eq:convex inequality}) can be actuated with sufficiently high pumping power and local regulation of the valves and any flow $\mathbf{q}$ that satisfies all of the $N$ equations on the form (\ref{eq:convex inequality}) is feasible, i.e. $\mathbf{q} \in \mathcal{Q}$. Note that the equation (\ref{eq:convex inequality}) is convex in $\mathbf{q}$ (assuming $c_1$ negative). Therefore the set $\mathcal{Q}$, as the union of $N$ convex sets on the form (\ref{eq:convex inequality}) and the constraints $\mathbf{q} \geq 0$, is also convex.
\begin{align}
    \hspace{-5pt}\mathcal{Q} &= \{ \q \hspace{1pt}| \hspace{1pt} \q \geq 0, \hspace{1pt} \q^T M_l \q - c_2 - c_3 \leq 0,\hspace{1pt} l = 1 \dots N  \} \\
    \hspace{-5pt}M_l &= D^T A D + E_{l,l}a_l^\text{min}
\end{align}
Here $D$ is an $N$ by $N$ upper triangular matrix of ones. A is an $N$ by $N$ diagonal matrix with entries $A_{1,1} = a_1^\text{sup} + a_1^\text{ret} - c_1$, $A_{i,i} = a_j^\text{sup} + a_j^\text{ret}$ for $2 \leq i\leq l$ and 0 otherwise. $E_{l,l}$ is an $N$ by $N$ matrix of all zeros, except for the element $l,l$, which is a one. This formulation makes $M_l$ an $N$ by $N$ positive semidefinite matrix since $A$ and $E_{l,l}$ have only positive diagonal entries, and thus $\mathcal{Q}$ is a union of quadratic and linear constraints, making it a convex set.
\\

\remark
The convexity of the set $\mathcal{Q}$ is connected to this specific grid structure, as the direction of the flow in this network is obvious. Indeed a meshed network is not guaranteed to enjoy this convexity of $\mathcal{Q}$, making optimization over the constraints on $\q$ harder to handle.
\\

\remark
Further operational constraints could also be introduced to restrict $\q$. For instance, too large flows may cause damage to pipes, or generate noise. An upper flow limitation could easily be added.

%%%%%%%%%%%%%%%%%%%%%%%%%%%%%%%%%%%%%%%%%%%%%%%%%%%%%%%%%%%%%%%%%%%%%%%%%%%%%%%%

\section{CONTROL STRATEGIES}\label{sec:control architectures}

This section investigates two potential control strategies, the \textit{traditional} strategy and the \textit{load coordination} strategy.  An \textit{optimal} baseline comparison is also introduced. The traditional architecture is where units are not connected through any sort of IT communication, and are simply attempting to maintain their own indoor temperature. In the load coordination architecture, the units calculate their desired loads locally through the exact same method as the traditional architecture, but these loads are then processed in a central computation and altered if they are not feasible. The optimal baseline is an upper bound on performance given the cost defined for the system. In this baseline it is assumed that a central unit has access to a perfect model of the entire system, as well as a posteriori measurements of the disturbance.
\\

\remark
Night set-back is an additional part of control strategies common in for instance Southern Europe. \cite{thebible} However, this work considers primarily the Northern European situation where this practice is less common and therefore it will not be considered.

\subsection{Traditional Architecture} \label{sec:traditional architecture}
In traditional DH systems there is no IT communication between units in the network. Each unit will greedily evaluate their own desired flow $q_i$ and actuate it through their control valve. The central pump then ensures that the pressure difference between supply and return pipes in the network is high enough to allow these control valves to actuate any desired flow. Traditionally, the control for each individual building has been done through the following control loop: A temperature curve is calibrated for the building, where a reference temperature $T_{\hs,i}^\text{r}$ is set for the water circulating in the heating system, $T_{\hs,i}$. A controller then tracks this reference through the control signal $q_i$, i.e. the flow through the substation heat exchanger primary side. This is actuated through altering the control valve opening $a_i(t)$. In this work we assume a simple proportional controller with gain $k_i$
\begin{align}
    T_{\hs,i}^\text{r} &= \alpha_{0,i} + \alpha_{1,i} T_\ext \label{eq:reference temperature} \\
    \Tilde{q}_i &= k_i(T_{\hs,i}^\text{r}-T_{\hs,i}) \label{eq:p controller}
\end{align}
Here $\alpha_{0,i}$ and $\alpha_{1,i}$ are calibration parameters for the temperature curve. $\Tilde{q}_i$ is the desired flow. When the distribution system is at maximum capacity, the differential pressure at unit $i$ may be too low, and in that case the actual flow $q_i$ will be lower than $\Tilde{q}_i$. The tuning of the parameters would be done by hand by a technician, based on experience and knowledge of suitable parameters for similar buildings. When a unit is not constrained in the flow $q_i(t)$, the unit should be able to reject the influence of outdoor temperatures such that a stationary outdoor temperature should not cause a stationary deviation in indoor temperature. When investigating the model of each building (\ref{eq:Tin}), (\ref{eq:Trad}) and (\ref{eq:p controller}), we can find that this is fulfilled when
\begin{align}
    1 + R_{\hs,i}\beta_i k_i + R_{\ext,i}\beta_i k_i \alpha_{1,i} &= 0, \label{eq:stationary condition 1}\\
    \frac{1}{1-\alpha_{1,i}} \alpha_{0,i} &= T_{c,i}. \label{eq:stationary condition 2}
\end{align}
The details of these relations are covered in Appendix 1. Parameters chosen in this way yield that the building will be able to reject the influence of outdoor temperature and maintain indoor temperature at comfort level. For simulation purposes, the parameters were chosen as
\begin{align}
    k_i &= \frac{T_c}{\alpha_{0,i}R_{\ext,i} - T_{c,i}R_{\hs,i} - T_{c,i} R_{\ext,i}} \label{eq:controller parameter k}\\
    \alpha_{1,i} &= -\frac{1 + k_i R_{\hs,i}}{k_i R_{\ext,i}} \label{eq:controller parameter a1}.
\end{align}
$\alpha_{0,i}$ is simply chosen large enough that the denominator of (\ref{eq:controller parameter k}) does not become negative.
\\

\remark
\review{In practice, the actuator in the building substation is the control valve, and current implementations of control systems may use this actuator directly to control the secondary side heating system temperature. In this case, the flow $q$ becomes an output of the system rather than an input. This problem is readily overcome through standard cascade control. In this setup, the flow $q$ will be the input that dictates the temperature of heating system water. This flow level will be the set-point for a secondary control loop where the valve position is used to actuate the desired flow. This adds the complexity of including the measurement of the flow into the control process. \cite{cascade}}

\subsection{Load Coordination Architecture} \label{sec:load coordination architecture}
The main contribution of this work is the proposition of the following control strategy: Each unit calculates their desired flow $\Tilde{q}_i$ as per the traditional strategy of section \ref{sec:traditional architecture}, equations (\ref{eq:reference temperature}) and (\ref{eq:p controller}). However, a central device ensures feasibility and fairness by providing each unit with an adjustment $\delta_i$ so that the actuated flow will be $q_i = \Tilde{q}_i + \delta_i$. In terms of IT communication and computational complexity, this method would be found between the traditional architecture and other optimization-based approaches. Depending on how $\delta_i$ is calculated, the central unit does not need access to internal building measurements, only their desired flow $\Tilde{q}_i$. The explicit models of building dynamics i.e. equations (\ref{eq:Tin}) and (\ref{eq:Trad}) are not needed in the central computation. Instead only the tuning parameters of the controllers can be utilized. The tuning for the controllers in each building can be done in a distributed fashion, so that a technician working on one individual unit does not affect the control of the whole system.
\\

The aim of the coordination is that the temperature deviations in each building should be distributed more fairly than in the non-coordinated traditional case. In Appendix 1, we show that given
\begin{itemize}
    \item the models of the buildings presented in section \ref{sec:buildings}, equations (\ref{eq:Tin}), (\ref{eq:Trad}) and (\ref{eq:p controller})
    \item and the local unit controllers from section \ref{sec:traditional architecture}, equations (\ref{eq:stationary condition 1}) and (\ref{eq:stationary condition 2}),
\end{itemize}
then given a constant temperature disturbance, each unit will converge to the following stationary indoor temperature deviation from comfort $e_i$:
\begin{equation}
    e_i = \frac{1}{k_i(1-\alpha_{1,i})}\delta_i
\end{equation}
While this stationary deviation fails to capture the time dynamics of the system, it is still a valuable metric. Should the system be subject to a constant outdoor temperature lower than the system is able to reject due to flow constraints, then the indoor temperature deviations will align with this distribution. This motivates the following coordination strategy:
\\

Define the parameters $\gamma_i$:
\begin{equation}
    \gamma_i = \frac{1}{k_i(1-\alpha_{1,i})}
\end{equation}
\review{The interpretation of this parameter is a weight provided to each building, indicating how much the deviation $\delta_i$ will affect them. Units with large controller gain parameters ($k_i$ and $\alpha_{1,i}$) will not be as impacted by the deviation term.} The coordination then wants to minimize the weighted indoor temperature deviations, which can be formulated as the following optimization problem:
\begin{align}
	&\underset{\delta}{\text{minimize}} && \max_i |\lambda_i \gamma_i \delta_i| \label{eq:load coordination cost}\\ 
	&\text{subject to}
	&& \mathbf{\Tilde{q}} - \mathbf{\delta} \in \mathcal{Q} \label{eq:load coordination constraints} 
\end{align}
$\mathcal{Q}$ is a union of quadratic constraints, and the objective function can be reformulated as a linear program. Therefore this becomes a quadratic program where the number of constraints and decision variables grows linearly with the number of connected units, making the problem readily solvable by standard quadratic program solvers. The actual cost to minimize is $J_\infty$ (\ref{eq:cost infnorm}). This is a simplified problem where instead the central coordinator minimizes the weighted stationary temperature that would arise from the coordination terms $\delta_i$. The weights $\lambda_i$ are design parameters that could be used to capture the quality of service requirements of specific units. For instance a hospital with harsh climate requirements may have a larger $\lambda_i$ than for example a residential building. In this work the influence of $\lambda_i$ will not be investigated, and thus we will from now on assume $\lambda_i=1$.
\\

\remark
Note that according to the current assumptions of individual unit controllers, this central coordination can be designed without explicit knowledge of the building parameters $R_{\ext,i}$, $R_{\hs,i}$, $C_{\iin ,i}$ or $C_{\hs ,i}$. The modelling effort is left to each individual unit in the form of controller tuning.

\subsection{Optimal Baseline} \label{sec:optimal architecture}
While we are interested in comparing the load coordination strategy to the traditional strategy, it is also interesting to see what the upper limit of optimality is. We consider the following problem
\begin{subequations}
\begin{alignat}{3}
&\!\min_{\mathbf{q}(t_k)}        &\qquad& \sum_{k=0}^K \underset{i}{\text{max}} (|e_i(t_k)t_s|) \label{eq:optProb}\\
&\text{subject to} &      & \mathbf{T}(t_{k+1}) = A\mathbf{T}(t_{k}) + B_q \mathbf{q}(t_k) + B_\ext T_\ext(t_k),\label{eq:constraint1}\\
&                  &      & \mathbf{q}(t_k) \in \mathcal{Q}.\label{eq:constraint2}\\
&                  &      & \mathbf{T}(t_0) = \mathbf{T}_0.\label{eq:constraint3}
\end{alignat}
\end{subequations}
which can directly be solved by optimization solvers, as the problem is convex. The problem implies minimizing the cost $J_\infty$ of equation (\ref{eq:cost infnorm}), subject to the dynamical constraints of the system. For larger networks and longer time-horizons, it will no longer be feasible to solve the entire problem at once as we have done here without adding computational power. \\

It should be clarified that this optimal baseline as explored in this paper is only presented as a point of reference for comparison with the other methods. In reality it would be completely unfeasible to have exact knowledge of all system parameters, system states, and knowledge of future disturbances. This comparison serves to give an indication about how much possible improvement a given strategy could theoretically have, given our current cost-evaluation.
\\

\remark
It should be noted that this is distinct from online optimization-and-prediction based strategies such as MPC. Such methods rely on online measurements and predictions of disturbances and state evolutions. The optimal strategy in this work is an a posteriori optimization given full knowledge of disturbances and system models.

%%%%%%%%%%%%%%%%%%%%%%%%%%%%%%%%%%%%%%%%%%%%%%%%%%%%%%%%%%%%%%%%%%%%%%%%%%%%%%%%

\section{Simulation and Results}\label{sec:simulation and results}
The first part of this section details the setup for the simulation experiments, followed by a part detailing the results.

\subsection{Simulation Description}

\begin{table*}[]
    \centering
    \caption{Model parameters used for simulation.}
    \resizebox{\textwidth}{!}{
    \begin{tabular}{c|lllllllllll} \toprule
        Index & a$^\text{ret}$  & a$^\text{sup}$ & a$^\text{min}$ & R$_\hs$ & R$_\ext$ & C$_\hs$ & C$_\iin$ & T$_c$ & $\alpha_0$ & $\alpha_1$ & k \\ \midrule
        Unit & mPa/(kg/s)$^2$ & mPa/(kg/s)$^2$ & mPa/(kg/s)$^2$ & mW/$\degC$ & mW/$\degC$ & kJ/$\degC$ & kJ/$\degC$ & $\degC$ & $\degC$ & - & kg/s$\degC$ \\ \midrule 
        \begin{filecontents*}{Data/parametertable.csv}
i,ssup,sret,umin,rhs,rext,chs,cin,tc,a0,a1,k,qmax
1,8.22,8.22,160,25.72,14.94,22.58,1109.99,20,65.3,-2.27,9.49,8.76
2,3.14,3.14,64.71,26.73,20.55,22.43,1017.96,20,55.22,-1.76,8.16,6.37
3,5.18,5.18,285.82,18.08,20.21,18.92,1050.65,20,45.47,-1.27,10.08,6.48
4,4.28,4.28,296.73,26.82,13.37,21.55,1269.63,20,72.14,-2.61,9.6,9.79
5,7.78,7.78,179.07,23.69,12.76,16.71,1183.97,20,68.55,-2.43,10.58,10.26
6,5.39,5.39,178.65,17.75,18.03,22.06,1111,20,47.62,-1.38,10.78,7.26
7,8.49,8.49,139.38,19.76,24.44,15.32,1398.5,20,43.4,-1.17,8.73,5.36
8,3.77,3.77,285.13,22.74,15.84,17.77,1251.16,20,58.46,-1.92,10,8.27
9,4.3,4.3,147.55,27.31,19.24,15.46,1229.83,20,58.06,-1.9,8.29,6.81
10,3.54,3.54,80.66,27.39,14.22,15.97,1450.32,20,70.23,-2.51,9.27,9.21
11,3.47,3.47,254.08,18.42,21.55,23.23,1071.5,20,44.52,-1.23,9.65,6.08
12,8.23,8.23,152.86,27.45,14.65,21.95,1354.32,20,68.96,-2.45,9.16,8.94
13,6.35,6.35,114.49,27.3,18.14,18.17,1352.24,20,60.12,-2.01,8.49,7.22
14,6.16,6.16,156.53,22.06,20.82,24.5,1128.27,20,49.43,-1.47,9,6.29
15,3.53,3.53,76.84,25.56,23.48,15.34,1240.69,20,50.12,-1.51,7.87,5.58
16,8.12,8.12,86.05,18.24,24.43,19.39,945.51,20,41.92,-1.1,9.04,5.36
17,6.62,6.62,296.02,21.35,18.71,18.82,932.37,20,51.39,-1.57,9.63,7
18,4.87,4.87,299.67,26.84,13.04,22.66,1218.48,20,73.41,-2.67,9.67,10.05
19,5.92,5.92,200.93,25.47,13.18,22.95,1367.5,20,70.36,-2.52,9.98,9.93
20,5.2,5.2,67.33,27.33,14.69,16.87,1460.41,20,68.65,-2.43,9.18,8.92
21,3.08,3.08,112.69,23.95,22.79,19.9,977.94,20,49.23,-1.46,8.25,5.75
22,4.15,4.15,143.38,17.06,14.64,19.46,1241.29,20,51.97,-1.6,12.17,8.94
23,3.39,3.39,264.69,26.1,22.42,21.46,1181.63,20,51.94,-1.6,7.95,5.84
24,3.78,3.78,55.83,27.04,14.49,22.09,907.14,20,68.78,-2.44,9.29,9.04
25,4.15,4.15,62.99,24.21,24.02,22.55,1102.27,20,48.19,-1.41,8,5.45
\end{filecontents*}
\csvreader[head to column names]{Data/parametertable.csv}{}
        {\csvcoli & \csvcolii & \csvcoliii & \csvcoliv & \csvcolv & \csvcolvi & \csvcolvii & \csvcolviii & \csvcolix & \csvcolx & \csvcolxi & \csvcolxii\\} \\ \bottomrule 
    \end{tabular}
    }
    \label{tab:network parameters}
\end{table*}

This work was simulated in Matlab, with optimization performed using Yalmip \cite{yalmip} with a Mosek optimizer. A network of $N = 25$ buildings, consisting of state space models as per section \ref{sec:buildings} was generated randomly. Controller parameters $k_i$, $\alpha_{0,i}$ and $\alpha_{1,i}$ were generated for each building in accordance with section \ref{sec:traditional architecture}. Random parameters were generated for pipes connecting these buildings in a line as per Figure \ref{fig:convex network}, as well as parameters that describe the limits of customer substations. The random generation of parameters was done by setting a nominal value for parameters based on parameters from similar models in other works, and then uniformly generating the parameters in a range from these nominal values. The resulting parameters are listed in Table \ref{tab:network parameters}
\\

The distribution pump curve (\ref{eq:pump curve}) was generated the following way: The pump is dimensioned to handle a peak load that occurs at -15$\degC$ outdoor temperature. For each building connected to the grid, the flow required to keep the unit at comfort temperature given an outdoor temperature of -15$\degC$, denoted $\q_i^\text{peak}$ was calculated, given equations (\ref{eq:Tin}) and (\ref{eq:Trad}). Using these flows in the left side of (\ref{eq:loop net pressure}) with $a_l(t) = a_l^\text{max}$, the pressure generated by the pump $p^\text{peak}$ can be calculated.
\begin{equation}
    p^\text{peak} = \max_l (q_l^{\text{peak}^2}a_l^{\text{min}} + \sum_{j=1}^l (a_j^\text{sup} + a_j^\text{ret})(\sum_{i=j}^N q_i^\text{peak})^2)
\end{equation}
It is then assumed that at this peak flow rate, the pump is running at full capacity, $r(t) = 1$. The parameters $c_i$ are then found by solving the equation
\begin{equation}
    c_1(\sum_{i=1}^N q_i^\text{peak})^2 + c_2 + c_3 = p^\text{peak}
\end{equation}
such that $c_i$ are proportional to the corresponding parameters in other literature \cite{hydraulic_optim_meshed}.
\\

\begin{figure}[]
    \centering
    \begin{tikzpicture}
				\begin{axis}[
					width=.8\columnwidth, height=5cm,
					xmin = 0, xmax = 200,
					ymin = -20, ymax = 5,
					grid = both,
					xlabel = {Time [h]},
					ylabel = {Outdoor Temperature [$\degC$]}
					]
					\addplot[blue] table[col sep = comma,x=time,y=Tout]{Data/Tout.csv};
				\end{axis}
			\end{tikzpicture}
    \caption{Outdoor temperature curve used for simulation.}
    \label{fig:outdoortemp}
\end{figure}
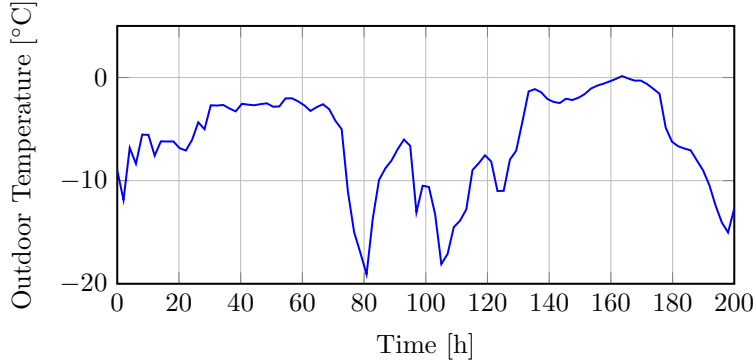

The system was then simulated subject to an outdoor temperature curve generated from real data. The data was gathered from \cite{smhi}, from a region in Sweden, chosen to represent a time period of drastically dropping temperature. The readings are hourly measurements and were therefore linearly interpolated to 15 minute intervals in the simulation. The resulting temperature curve is visible in Figure \ref{fig:outdoortemp}. The simulation was done for each of the above listed architectures.

\subsection{Results}

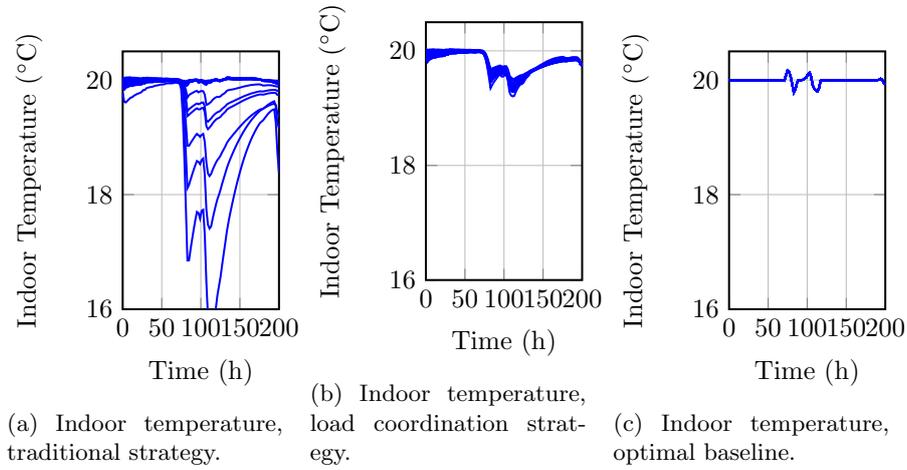
\begin{figure*}[]
    \centering
    \begin{subfigure}[b]{.3\textwidth}
        \begin{tikzpicture}
            \begin{axis}[
                width=\columnwidth, height=5cm,
                xmin = 0, xmax = 200,
                ymin = 16, ymax = 20.5,
                grid = both,
                xlabel = Time (h),
                ylabel = Indoor Temperature ($\degC$)
                ]
                \foreach \i in {1,...,25}
                \addplot[blue] table[col sep = comma,x=time,y=\i]{Data/Tin_g.csv};
            \end{axis}
        \end{tikzpicture}
        \caption{Indoor temperature, traditional strategy.}
        \label{fig:traditional Tin}
    \end{subfigure}
    ~
    \begin{subfigure}[b]{.3\textwidth}
        \begin{tikzpicture}
            \begin{axis}[
                width=\columnwidth, height=5cm,
                xmin = 0, xmax = 200,
                ymin = 16, ymax = 20.5,
                grid = both,
                xlabel = Time (h),
                ylabel = Indoor Temperature ($\degC$)
                ]
                \foreach \i in {1,...,25}
                \addplot[blue] table[col sep = comma,x=time,y=\i]{Data/Tin_c.csv};
            \end{axis}
        \end{tikzpicture}
        \caption{Indoor temperature, load coordination strategy.}
        \label{fig:coordinated Tin}
    \end{subfigure}
    ~
    \begin{subfigure}[b]{.3\textwidth}
        \begin{tikzpicture}
            \begin{axis}[
                width=\columnwidth, height=5cm,
                xmin = 0, xmax = 200,
                ymin = 16, ymax = 20.5,
                grid = both,
                xlabel = Time (h),
                ylabel = Indoor Temperature ($\degC$)
                ]
                \foreach \i in {1,...,25}
                \addplot[blue] table[col sep = comma,x=time,y=\i]{Data/Tin_o.csv};
            \end{axis}
        \end{tikzpicture}
        \caption{Indoor temperature, optimal baseline.}
        \label{fig:optimal Tin}
    \end{subfigure}
    
    \caption{Indoor temperatures registered during the simulation.}
    \label{fig:simulation temperatures}
\end{figure*}

Figures \ref{fig:traditional Tin}, \ref{fig:coordinated Tin} and \ref{fig:optimal Tin} show the evolution of indoor temperatures using the traditional strategy, load coordination strategy and optimal baseline respectively. Recall from the problem formulation of section \ref{sec:problem formulation} that no unit should experience heavy temperature deviations from the comfort temperature of 20$\degC$. The clear distinction between the strategies is that using the traditional architecture results in a few units deviating greatly from their desired indoor temperature. Using the load coordination strategy, the units are much more aligned, leading to all units experiencing deviations but on a much lower magnitude. Finally, in the optimal baseline the results are even better. The units hardly deviate at all from their desired temperatures, and temperatures are deviating equally between all units. In this baseline the units are also pre-heated before the severe drop in temperature, which is not incorporated in the other strategies as they do not include any predictive behaviour. 
\\

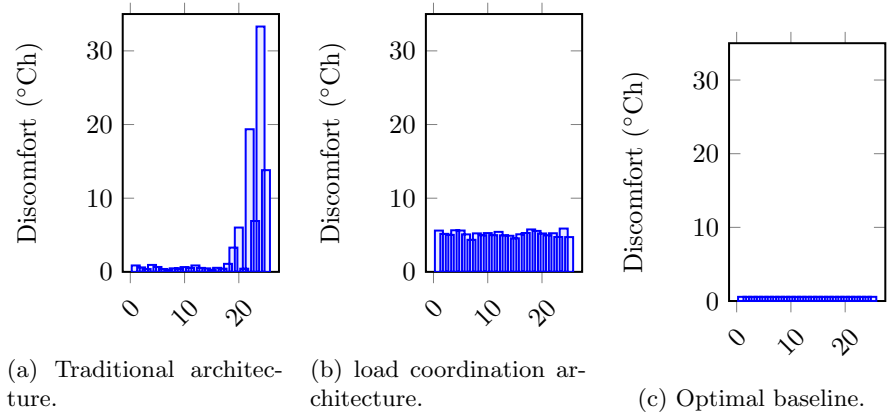
\begin{figure*}
    \centering
    \begin{subfigure}[b]{.3\textwidth}
        \begin{tikzpicture}
            \begin{axis}[
                width=\columnwidth, height=5cm,
                ybar,
                ymin = 0, ymax = 35,
            	ylabel=Discomfort ($\degC$h),
            	x tick label style={rotate=45},
            	bar width = 3pt,
            ]
            \addplot [blue, fill = blue!8] table[col sep = comma, x=unit, y=greedy]{Data/discomforts.csv};
            \end{axis}
        \end{tikzpicture}
        \caption{Traditional architecture.}
        \label{fig:discomfort traditional}
    \end{subfigure}
    ~
    \begin{subfigure}[b]{.3\textwidth}
        \begin{tikzpicture}
            \begin{axis}[
                width=\columnwidth, height=5cm,
                ybar,
                ymin = 0, ymax = 35,
            	ylabel=Discomfort ($\degC$h),
            	x tick label style={rotate=45},
            	bar width = 3pt,
            ]
            \addplot [blue, fill = blue!8] table[col sep = comma, x=unit, y=controller]{Data/discomforts.csv};
            \end{axis}
        \end{tikzpicture}
        \caption{load coordination architecture.}
        \label{fig:discomfort coordinated}
    \end{subfigure}
    ~
    \begin{subfigure}[b]{.3\textwidth}
        \begin{tikzpicture}
            \begin{axis}[
                width=\columnwidth, height=5cm,
                ybar,
                ymin = 0, ymax = 35,
            	ylabel=Discomfort ($\degC$h),
            	x tick label style={rotate=45},
            	bar width = 3pt,
            ]
            \addplot [blue, fill = blue!8] table[col sep = comma, x=unit, y=optimal]{Data/discomforts.csv};
            \end{axis}
        \end{tikzpicture}
        \caption{Optimal baseline.}
        \label{fig:discomfort optimal}
    \end{subfigure}
    
    \caption{Discomfort experienced by each individual building, indexed 1-25 by their distance from the central distribution pump where 25 is the furthest.}
    \label{fig:simulation discomfort}
\end{figure*}

The plots of Figures \ref{fig:simulation temperatures} give a hint of what the effect of the different strategies are. However they are also supported by Figure \ref{fig:simulation discomfort}. Here the discomfort metric of equation (\ref{eq:discomfort}) are shown, evaluated on each strategy and unit. Figure \ref{fig:discomfort traditional} shows the inequality generated by the traditional strategy, as units located further from the heat source experience higher discomfort. Meanwhile, Figure \ref{fig:discomfort coordinated} shows a much more equal distribution of discomfort. Lastly, Figure \ref{fig:discomfort optimal} shows that there is still a discrepancy between the coordinated strategy and the theoretical lower bound on discomfort.
\\

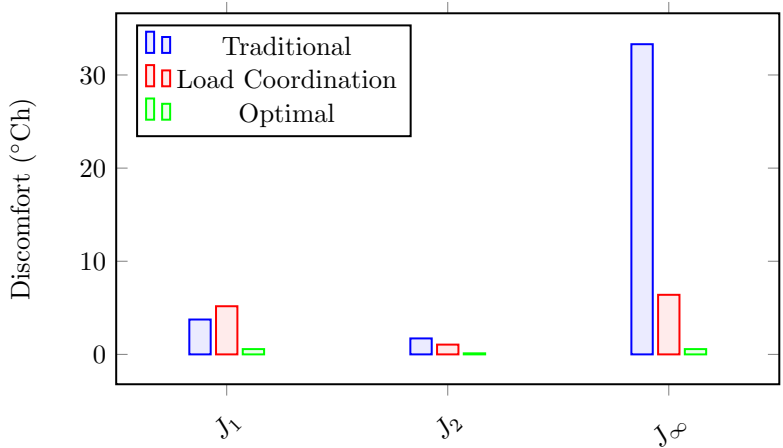
\begin{figure}[]
    \centering
    \begin{tikzpicture}
    \begin{axis}[
        width=.85\linewidth, height=6.5cm,
        ybar,
        bar width = 8pt,
    	ylabel=Discomfort ($\degC$h),
    	xtick = {1,2,3},
    	xticklabels = { $J_1$, $J_2$, $J_\infty$},
    	legend pos = north west,
    	x tick label style={rotate=45},
        xmin = .5, xmax = 3.5
    ]
    \addplot [blue, fill = blue!8] table[col sep = comma, x = tick, y=greedy]{Data/j_costs_cropped.csv};
    %\addplot [orange, fill = orange!8] table[col sep = comma, x = tick, y=naive]{Data/j_costs.csv};
    \addplot [red, fill = red!8] table[col sep = comma, x = tick, y=controller]{Data/j_costs_cropped.csv};
    \addplot [green, fill = green!8] table[col sep = comma, x = tick, y=optimal]{Data/j_costs_cropped.csv};
    \legend{Traditional, Load Coordination, Optimal}
    \end{axis}
    \end{tikzpicture}
    \caption{Discomfort metrics defined in equations (\ref{eq:cost 1norm}), (\ref{eq:cost 2norm}) and (\ref{eq:cost infnorm}) ($J_1$, $J_2$ and $J_\infty$ respectively) evaluated through each coordination strategy.}
    \label{fig:cost comparison}
\end{figure}

Figure \ref{fig:cost comparison} shows the different discomfort metrics of (\ref{eq:cost 1norm}), (\ref{eq:cost 2norm}) and (\ref{eq:cost infnorm}) evaluated through each coordination strategy, corresponding to $J_1$, $J_2$ and $J_\infty$ respectively. We see that the sum of discomfort experienced in units, corresponding to $J_1$, is actually improved using traditional architecture than the load coordination architecture. This is quite reasonable, since providing higher flow to units further down the network incurs a higher pressure loss. Thus the total flow provided in the traditional strategy is higher. However, when measured through $J_2$ and $J_\infty$, the load coordination strategy outperforms the traditional strategy. This is because the worst-case experience for any unit is much lower with this setup. The optimal baseline shows that there is still potential improvements to be made.
\\

%%%%%%%%%%%%%%%%%%%%%%%%%%%%%%%%%%%%%%%%%%%%%%%%%%%%%%%%%%%%%%%%%%%%%%%%%%%%%%%%

\section{Summary}\label{sec:summary}
This section concludes the work with some final remarks, followed by potential future outlooks.

\subsection{Conclusions}
In this work, we investigated the influence of two different architectures for coordinating the flows in a line-structured district heating network. It was shown that utilizing traditional control strategies in each unit can be augmented with a coordination mechanism which reduces the worst-case discomfort experienced by any unit under peak load conditions, at the cost of increasing the mean discomfort, see \ref{fig:cost comparison}. This coordination can be achieved without explicit models or temperature readings accessed by the central unit. \review{This proof of concept shows how augmenting future district heating systems with smarter controllers can increase the systems' robustness to peak load conditions. The design requirements for future district heating grids can therefore be lowered, allowing for lower grid temperature without as much additional grid capacity in terms of extended piping and pumping power.} 
\\

However, further improvements can be made to the control strategy when utilizing an optimization-based architecture that allows utilizing information on temperature forecasts to pre-heat units ahead of peak loads. This requires even further complexity, where the central computation unit would have access to individual unit measurements, unit building parameters, and accurate weather forecasts.
\\

The fact that the coordination strategy does not rely on building temperature measurements, and that controllers can be tuned individually for units without affecting the tuning of other units, makes the strategy scalable to growing networks as well as a more privacy-compliant option than a full optimization-based scheme.

\subsection{Future Work}

The proposed coordination strategy currently does not include the intelligent behavior of the optimal baseline, where the unit indoor temperatures can be utilized for pre-heating before load peaks, often referred to as peak-shaving and valley-filling. The main interest here would be to see if the architecture could maintain the autonomy of unit controllers, while simultaneously including predictive behavior based on an outdoor temperature forecast.
\\

Both the optimal strategy and the proposed coordination strategy currently rely on understanding the set $\mathcal{Q}$ that describes the set of possible flows. This may in practice be harder estimate than proposed in this work, as specific and accurate parameters for all network parameters may not be known, or degrade and change over time. Therefore it would be interesting to see how these methods hold to uncertainties in network models, as well as data driven methods for estimating the parameters that dictate $\mathcal{Q}$. \review{While the building model parameters are technically not necessary in the controller coordination, it is reasonable to believe that building controllers will not be as perfectly tuned as proposed in this work. Therefore a study should be conducted to investigate the sensitivity to poorly tuned individual building controllers.}
\\

To further simplify the tuning of individual unit controllers, it is likely that more sophisticated unit controllers should be utilized. For instance, a simple PI-controller would allow the elimination of stationary errors when tracking the reference heating system temperature. Therefore including more advanced individual controllers in the analysis would be a valuable extension.

\bibliographystyle{unsrt}
\bibliography{bibliography}

\begin{thebibliography}{10}

\bibitem{smartenergyfor100renewable}
B.V. Mathiesen, H.~Lund, D.~Connolly, H.~Wenzel, P.A. Østergaard, B.~Möller,
  S.~Nielsen, I.~Ridjan, P.~Karnøe, K.~Sperling, and F.K. Hvelplund.
\newblock Smart energy systems for coherent 100\% renewable energy and
  transport solutions.
\newblock {\em Applied Energy}, 145:139--154, 2015.

\bibitem{lund_4th_2014}
H.~Lund, S.~Werner, R.~Wiltshire, S.~Svendsen, J.E. Thorsen, F.~Hvelplund, and
  B.V. Mathiesen.
\newblock 4th {generation} {district} {heating} ({4GDH}).
\newblock {\em Energy}, 68:1--11, April 2014.

\bibitem{statusof4g}
H.~Lund, P.~A. Østergaard, M.~Chang, S.~Werner, S.~Svendsen, P.~Sorknæs,
  J.~E. Thorsen, F.~Hvelplund, B.~O.~G. Mortensen, B.~V. Mathiesen, C.~Bojesen,
  N.~Duic, X.~Zhang, and B.~Möller.
\newblock The status of 4th generation district heating: Research and results.
\newblock {\em Energy}, 164:147--159, 2018.

\bibitem{comparisonoflowtemp}
R.~S. Lund, D.~S. {\O}stergaard, X.~Yang, and B.~V. Mathiesen.
\newblock Comparison of low-temperature district heating concepts in a
  long-term energy system perspective.
\newblock {\em International Journal of Sustainable Energy Planning and
  Management}, 12:5--18, 2017.

\bibitem{brange_decision-making_2019}
L.~Brange, K.~Sernhed, and M.~Thern.
\newblock Decision-making process for addressing bottleneck problems in
  district heating networks.
\newblock {\em International Journal of Sustainable Energy Planning and
  Management}, 20:37--50, 2019.

\bibitem{thebible}
S.~Frederiksen and S.~Werner.
\newblock {\em District heating and cooling}.
\newblock Studentlitteratur, 2013.

\bibitem{vandermeulen_controlling_2018}
A.~Vandermeulen, B.~van~der Heijde, and L.~Helsen.
\newblock Controlling district heating and cooling networks to unlock
  flexibility: {A} review.
\newblock {\em Energy}, 151:103--115, May 2018.

\bibitem{guelpa_demand_2021}
E.~Guelpa and V.~Verda.
\newblock Demand response and other demand side management techniques for
  district heating: {A} review.
\newblock {\em Energy}, 219, March 2021.

\bibitem{bhattacharya_demand_2019}
S.~Bhattacharya, Chandan V., Arya V., and Kar K.
\newblock Demand {response} for {thermal} {fairness} in {district} {heating}
  {networks}.
\newblock {\em IEEE Transactions on Sustainable Energy}, 10(2):865--875, April
  2019.

\bibitem{saletti_development_2020}
C.~Saletti, A.~Gambarotta, and M.~Morini.
\newblock Development, analysis and application of a predictive controller to a
  small-scale district heating system.
\newblock {\em Applied Thermal Engineering}, 165, January 2020.

\bibitem{sarbu_review_2019}
I.~Sarbu, M.~Mirza, and E.~Crasmareanu.
\newblock A review of modelling and optimisation techniques for district
  heating systems.
\newblock {\em International Journal of Energy Research}, pages 6572--6598, May
  2019.

\bibitem{hydraulic_optim_meshed}
Y.~Wang, S.~You, H.~Zhang, W.~Zheng, X.~Zheng, and Q.~Miao.
\newblock Hydraulic performance optimization of meshed district heating network
  with multiple heat sources.
\newblock {\em Energy}, 126:603--621, 03 2017.

\bibitem{benny_operational_nodate}
A.~Benonysson, B.~Bøhm, and H.F. Ravn.
\newblock Operational optimization in a district heating system.
\newblock {\em Energy Conversion and Management}, 36:297--314, 1995.

\bibitem{larsen_aggregated_2002}
H.V. Larsen, H.~Pálsson, B.~Bøhm, and H.F. Ravn.
\newblock Aggregated dynamic simulation model of district heating networks.
\newblock {\em Energy Conversion and Management}, 43(8):995--1019, May 2002.

\bibitem{bacher_identifying_2011}
P.~Bacher and H.~Madsen.
\newblock Identifying suitable models for the heat dynamics of buildings.
\newblock {\em Energy and Buildings}, 43(7):1511--1522, July 2011.

\bibitem{SULZERPUMPS201027}
Sulzer Pumps.
\newblock {\em Centrifugal pump handbook}.
\newblock Butterworth-Heinemann, Oxford, 3rd edition, 2010.

\bibitem{cascade}
S.~Skogestad and I.~Postlethwaite.
\newblock {\em Multivariable Feedback Control: Analysis and Design}.
\newblock John Wiley \& Sons Ltd, Chichester, 2nd edition, 2005.

\bibitem{yalmip}
J.~L{\"{o}}fberg.
\newblock Yalmip : A toolbox for modeling and optimization in matlab.
\newblock In {\em In Proceedings of the CACSD Conference}, Taipei, Taiwan,
  2004.

\bibitem{smhi}
Swedish Meteorological and Hydrological Institute.
\newblock
  https://www.smhi.se/en/services/open-data/search-smhi-s-open-data-1.81004.

\end{thebibliography}

\section*{APPENDIX I - Individual Unit Controller Tuning}\label{sec:appendix 1}

This appendix presents the motivating equations behind the tuning of traditional unit controllers and the choice of weights $\gamma_i$ used in the load coordinating architecture.

To investigate the effects of the control parameters $k_i$, $\alpha_{0,i}$, $\alpha_{1,i}$ as well as the coordination signal $\delta_i$, we combine equations (\ref{eq:Tin}), (\ref{eq:Trad}), (\ref{eq:P}) and (\ref{eq:p controller}). This grants the following system description of the unit:

\begin{equation}
    \begin{pmatrix}
    \dot{T}_{\iin, i}\\
    \dot{T}_{\rad, i}
    \end{pmatrix} = 
    A \begin{pmatrix}
    T_{\iin, i}\\
    T_{\rad, i}
    \end{pmatrix} + B\begin{pmatrix}
    T_\ext \\
    \delta_i
    \end{pmatrix} + C
\end{equation}

where
\begin{equation}
    A = \begin{pmatrix}
    -\frac{1}{R_{\ext ,i}}-\frac{1}{R_{\rad ,i}} & \frac{1}{R_{\rad ,i}} \\
    \frac{1}{R_{\rad ,i}} & - \frac{1}{R_{\rad ,i}} - C_w (T_\text{sup} - T_\text{ret})k_i
    \end{pmatrix}
    \label{eq:A matrix}
\end{equation}

\begin{equation}
    B = \begin{pmatrix}
    \frac{1}{R_{\ext,i}} & 0\\
     C_w (T_\text{sup} - T_\text{ret})k_i\alpha_{1,i} & C_w (T_\text{sup} - T_\text{ret})
    \end{pmatrix}
    \label{eq:B matrix}
\end{equation}
and
\begin{equation}
    C = \begin{pmatrix}
    0 \\
    C_w (T_\text{sup} - T_\text{ret})k_i\alpha_{0,i}
    \end{pmatrix}
    \label{eq:C matrix}
\end{equation}

Note that these matrices $A$ and $B$ are not the same matrices as in equation (\ref{eq:temp_dynamics}). A feasible target for the design of the control parameters $k_i$, $\alpha_{0,i}$ and $\alpha_{1,i}$ is that when there is no coordination signal $\delta_i$, the building should, given a constant outdoor temperature $T_\ext^0$, be able to reach a given comfort temperature $T_{c,i}$ indoors. We therefore investigate the stationary case where $\dot{T}_{\iin, i} = \dot{T}_{\rad, i} = 0$, $T_\ext = T_\ext^0$ and $\delta_i = \delta_i^0$. We can find the resulting indoor and heating system temperatures as 
\begin{equation}
    \begin{pmatrix}
    T_{\iin, i}^0\\
    T_{\rad, i}^0
    \end{pmatrix} = 
    -A^{-1}B\begin{pmatrix}
    T_\ext^0 \\
    \delta_i^0
    \end{pmatrix} -A^{-1}C.
    \label{eq:stationary matrix equation}
\end{equation}

Introducing $\beta_i = C_w (T_\text{sup} - T_\text{ret})$ for brevity, this yields the following stationary indoor temperature:

\begin{align}
    T_{\iin, i}^0   &= \frac{1 + R_{\hs,i}\beta_i k_i + R_{\ext,i}\beta_i k_i \alpha_{1,i}}{1+R_{\ext,i}\beta_i k_i + R_{\hs,i}\beta_i k_i} T_\ext^0\label{eq:Text stationary term}\\
                    &+ \frac{R_{\ext,i}\beta_i}{1+R_{\ext,i}\beta_i k_i + R_{\hs,i}\beta_i k_i} \delta^0 \label{eq:delta stationary term}\\
                    &+ \frac{R_{\ext,i}\beta_i k_i}{1+R_{\ext,i}\beta_i k_i + R_{\hs,i}\beta_i k_i} \alpha_{0,i}\label{eq:a0 stationary term}
\end{align}

The temperature deviation caused by the external temperature is captured in the term (\ref{eq:Text stationary term}). To ensure that the outdoor temperature does not cause systematic temperature deviations, the controller gains will have to be chosen so that

\begin{equation}
    1 + R_{\hs,i}\beta_i k_i + R_{\ext,i}\beta_i k_i \alpha_{1,i} = 0
    \label{eq:choosing k and a1}
\end{equation}

Substituting equation (\ref{eq:choosing k and a1}) into the terms (\ref{eq:Text stationary term}), (\ref{eq:delta stationary term}) and (\ref{eq:a0 stationary term}), we receive the following resulting indoor temperature:

\begin{align}
    T_{\iin, i}^0   &=  \frac{1}{k_i(1-\alpha_{1,i})} \delta^0 \label{eq:delta final term}\\
                    &+ \frac{1}{1-\alpha_{1,i}} \alpha_{0,i}\label{eq:a0 final term}
\end{align}
From here, we see that a suitable choice of $\alpha_{0,i}$ is so that the relation

\begin{equation}
    \frac{1}{1-\alpha_{1,i}} \alpha_{0,i} = T_{c,i}
\end{equation}
is fulfilled, i.e. given no coordination term $\delta_i$, the unit should experience comfort temperature.

The remaining deviation caused by the coordination term $\delta_i$ is demonstrated in equation (\ref{eq:delta final term}), motivating the weights chosen in section \ref{sec:load coordination architecture}.

\end{document}